\documentclass{ws-procs9x6}

\newcommand\beq{\begin{equation}}
\newcommand\eeq{\end{equation}}

\begin{document}

\title{MAGNETIC STRINGS AS PART OF YANG--MILLS PLASMA}

\author{M.\,N.\,CHERNODUB}
\address{
Institute for Theoretical and Experimental Physics,\\
B. Cheremushkinskaya 25, 117118 Moscow, Russia\\
Research Institute for Information Science and Education,\\
Hiroshima University, Higashi-Hiroshima, 739-8527, Japan}

\author{V.\,I.\,ZAKHAROV}
\address{
Istituto Nazionale di Fisica Nucleare - Sezione di Pisa,\\
Largo Pontecorvo, 3, 56127 Pisa, Italy\\
Max-Planck Institut f\"ur Physik, F\"ohringer Ring 6, 80805, M\"unchen,
Germany}

\begin{abstract}
Magnetic strings are defined as infinitely thin surfaces which are closed in
the vacuum and can be open on an external monopole 
trajectory (that is, defined by 
't Hooft loop). We review briefly lattice data on the magnetic strings
which refer mostly to SU(2) and SU(3) pure Yang-Mills theories
and concentrate on implications of the strings for the Yang-Mills plasma. 
We argue that magnetic strings might be a liquid component of
the Yang-Mills plasma and suggest tests of this hypothesis.
\end{abstract}


\bodymatter

\section{Generalities}
\subsection{Definition of magnetic strings}

There are two fundamental probes of Yang-Mills vacuum, Wilson and 
't Hooft loops, related to the heavy-quark and heavy-monopole potentials:
\begin{eqnarray}
\langle W\rangle \sim \exp{(-V_{Q\bar{Q}}(r)\, T)}~,~~~~~~~~
\langle H\rangle \sim \exp{(-V_{M\bar{M}}(r)\, T)}\,.
\end{eqnarray}
In the string picture, 
one postulates that the Wilson and 't Hooft loops
are given by sums over all surfaces which can be spanned on the loops,
\begin{eqnarray}\label{open}
\langle W\rangle \sim
\sum
\exp(~-f(...)A_{C})\,,\\ \nonumber
\langle H\rangle \sim
\sum
\exp(~-\tilde{f}(...)A_{H})\,,
\end{eqnarray}
where $A_{C}$ and $A_{H}$ are the area of the surfaces spanned 
on the Wilson contour $C$ and the 't Hooft loop,
respectively, and $f(...)$, $\tilde{f}(...)$ are certain weight functions.
We do not specify the arguments of the weight functions here.
In models with extra dimensions, for example, the weight functions
depend on the background metrics.

Equations (\ref{open}) imply that there exist strings which can be open on the
Wilson and 't Hooft 
loops.
We call these strings as electric and magnetic strings, respectively.
Explicit realization of (\ref{open}) is achieved in
dual formulations of highly symmetrical versions of Yang-Mills theories,
for a review see Ref.~\refcite{klebanov}. 

\subsection{Percolation of strings}

At zero temperature, heavy quarks are confined while heavy monopoles are not
confined. In terms of the string tensions this means:
\begin{eqnarray}\label{vanishings}
\sigma_{\mathrm{electric}}(T=0) & \sim & \Lambda_{\mathrm{QCD}}^{2}\,,\\ \nonumber
\sigma_{\mathrm{magnetic}}(T=0)   &   =  & 0\,.
\end{eqnarray}
These relations are realized explicitly in case of large number
of colors $N_{c}$  and supersymmetry \cite{klebanov}.
A reservation is that the tension depends actually 
on the length of the strings and Eqs. (\ref{vanishings}) hold only if the 
strings are long enough.

Connection between the weight factors which enter (\ref{open}) and
the string tensions (\ref{vanishings}) is not so straightforward.
Consider first the case when 
the weight factors $f,\tilde{f}$ are constants. Then they still do not
coincide with the string tensions. 
The point is that the weight factors specify suppression due
to the action associated with the surface. However, there is a
hidden entropy factor in Eq.  (\ref{open}) which is the number of
various surfaces of the same area. 
Thus, the tension (\ref{vanishings}) reflects the balance
between suppression due to the action and enhancement due to the
entropy:
\begin{equation}
(\mbox{tension})\cdot (\mbox{area})~~\sim~~(\mbox{action})~-~(\mbox{entropy})\,.
\end{equation} 
One can estimate readily that the entropy factor results in exponential
enhancement of contribution of (large) areas. 

One of central points is that vanishing string tension does not mean
that the strings are not relevant and uninteresting. In a way, just
to the contrary. Vanishing tension means percolation of strings,
or that they 'are everywhere'. For the first time this observation was made 
in connection with deconfinement transition in Yang--Mills theories
at finite temperature \cite{polyakov}.
Namely, it was suggested that at the critical temperature the electric
strings become tensionless,
\begin{equation}
\sigma_{\mathrm{electric}}(T=T_c) = 0\,,
\end{equation}
and percolate through the vacuum.
 
  It is worth emphasizing that percolation assumes cancellation
between energy and entropy for long strings\footnote{Short strings
which quantum mechanically correspond to glueballs of low spin
can well have finite tension and, respectively, glueballs remain
massive at the phase transition.}. In particular, 
strong cancellation between energy and entropy of strings at temperatures
above the phase transition was observed in Ref.~\refcite{kaczmarek}
by studying the Wilson loop:
\begin{equation}
\langle W\rangle \sim \exp(-F)\,, \qquad F~=~U~+~T(\partial F/\partial T)\,,
\end{equation}
where $F$ is the free energy, $U$ is the potential energy.
Measured values of energy and entropy are about 10 and cancel each other.

Let us go back to discuss the $T=0$ case. Vanishing 
tension of magnetic strings, see (\ref{vanishings}) implies percolation
of magnetic strings at zero temperature. 
And, indeed, percolation of magnetic strings has been observed
on the lattice. In lattice terminology, magnetic strings are known 
as center vortices whose role in confinement has been extensively studied~\cite{greensite}.
 
  We are more interested in the structure of the strings themselves.
 In particular, if the strings are infinitely thin indeed then both energy and
 entropy should be divergent at short distances and cancel each other
 in free energy, or tension\footnote{For further comments and explanation see Ref.~\refcite{vz}.}. 
 This remarkable phenomenon was indeed observed \cite{ft}.
 In particular the non-Abelian action associated with the surfaces
 equals to:
 \begin{equation}\label{spacing}
 S_{\mathrm{strings}}~\approx~0.54 \, {(\mbox{Area}) \, a^{-2}}~~,
 \end{equation}
 where $a$ is the lattice spacing, $a\to 0$ in the continuum limit, and
 $(Area)$ is the total area which is in physical units. The data refers
 to $a\ge 0.06\,\mbox{fm}$.

\subsection{Supercritical phase}

The tachyonic nature of the magnetic strings is manifested
in existence of an infinite cluster of surfaces in the vacuum\footnote{For 
introduction to the percolation theory and  further references see, e.g., Ref.~\refcite{grimmelt}.}.
Indeed, the spatial extension of cluster is obviously related
to correlation length,
\begin{equation}
\langle R_{\mbox{finite cluster}}\rangle~\sim~l_{\mathrm{corr}}~\sim~m^{-1}\,,
\end{equation}
where $m$ is the excitation mass. With a stretch of imagination,
one can conclude that existence of an infinite
cluster corresponds then either to zero or tachyonic mass\footnote{The absolute value 
of the tachyonic mass controls in fact density of the infinite cluster.}.

The most  non-trivial ingredient of  the percolation theory
is that there is continuity of description across the threshold
of emergence of a tachyonic mode. In other words, one
can still use the language of the false vacuum even 
when there exists a negative mode. An example of 
such a description is an expression for the
density of the infinite cluster as proportional to $\zeta^{\alpha}$ where $\zeta=0$ at the
point of the phase transition, where $\zeta$ is a non-negative function of couplings of the model
and $\alpha$ is a positive number.

The density of the percolating cluster of the magnetic strings does
satisfy such a relation:
\begin{equation}
\theta_{\mathrm{plaq}}~\approx~ \biggl({\frac{a}{0.5\,\mbox{fm}}}\biggr)^{2}~~,
\end{equation}
where $\theta_{\mathrm{plaq}}$ is the probability for a given plaquette
on the lattice to belong to the infinite, or percolating cluster of
surfaces. Note that in the continuum limit of $a\to 0$ the infinite cluster
occupies a vanishing part of the total volume. The value of
$(a\cdot\Lambda_{\mathrm{QCD}})$ specifies closeness to the point of 
the phase transition.

Apart from the probability $\theta_{\mathrm{plaq}}$ one can introduce
many other observables and corresponding critical exponents.
Theory of percolation reduces then to relations among the critical
exponents.

\subsection{Magnetic monopoles}

Theory of percolation of surfaces is much less developed than theory
of percolation of trajectories (respectively, of strings and of particles).
Thus, it is a great simplification that  magnetic strings
can in fact be substituted for our purposes by particles living
on the strings. In lattice terminology\footnote{The terminology is somewhat misleading since
the non-Abelian field of the monopole--like objects considered here
is aligned with the surfaces and in general is not quantized.
The field of conventional ``monopoles'' in a common sense of this word would
be, instead, spherically symmetric and quantized.} the particles are 
magnetic monopoles, for further references see Ref.~\refcite{vz}.
Geometrically the substitution is possible since the monopole
trajectories (which are closed loops) cover densely the surfaces
associated with the strings. Physicswise, the relation is that monopoles
correspond to a tachyonic mode in terms of the string excitations.

The probability of a given link to belong to a monopole trajectory is\cite{bornyakov}
\begin{equation}\label{density}
\theta_{\mathrm{link}}~\approx~\biggl(\frac{a}{0.5\,\mbox{fm}}\biggr)^{3}~~
+~\biggl(\frac{a}{0.8\,\mbox{fm}}\biggr)^{2}\biggl(1~-~\frac{a}{0.25\,\mbox{fm}}\biggr)\,,
\end{equation}
where $a< 0.13\,\mbox{fm}$ for the theory to converge to the continuum limit.
The first term here corresponds to the infinite cluster, or the tachyonic mode.
The second  term corresponds to finite clusters, or excitations
and dominates in the limit of $a\to 0$. 
Moreover, numerically about (90-95)\% of monopole trajectories
fall on the  magnetic-strings surfaces, see Ref.~\refcite{ft} and references therein. 
This is just clarification what we meant above by ``particles living on the
strings'': the evidence is pure numerical but convincing.

\section{Magnetic component of Yang-Mills plasma}
\subsection{From tachyonic mode to thermal plasma}

Percolating magnetic strings is a fascinating phenomenon. There is a
word of caution, however: magnetic strings describe vacuum structure,
or the structure of the ground state. What is actually observable are
excitations of the ground state.  The percolating vacuum magnetic strings 
represent  a tachyonic mode with respect to the perturbative vacuum
and cannot be observed directly in an experimental set-up.
Things may change at finite temperature, as we will explain 
now.

Indeed, it is a general rule that degrees of freedom which are virtual
at zero temperature become real particles at non-zero temperature.
Consider, for example, theory of free photons. At zero temperature
the lowest state, vacuum is realized as zero-point fluctuations.
If we evaluate the energy density it is ultraviolet divergent:
$$
%stilisticheskie izmenenija
\epsilon_{\mathrm{vac}}~\sim~\sum \frac{\omega ({\bf k})}{2}
\sim \Lambda_{\mathrm{UV}}^{4},$$
where ${\bf k}$ is the momentum
and $\Lambda_{\mathrm{UV}}$ is an ultraviolet cut off. At finite temperature,
the energy density of the photon gas is given by the Stefan-Boltzmann law
$$\epsilon(T)~\sim~T^{4}\,,$$
and, obviously, the coefficients in front of the ultraviolet divergence at zero
temperature and in front of $T^{4}$ are related to each other.

An example closer to our case: if there is Higgs mechanism 
of spontaneous symmetry breaking at zero temperature,
at finite temperature the symmetry is restored and the number
of degrees of freedom of scalar particles in plasma is the same as in the Lagrangian
(while at zero temperature the tachyonic mode is unobservable).

Thus we can expect, that at deconfinement phase transition 
magnetic strings, or monopoles are released into the thermal plasma~\cite{chernodub}.

\subsection{Percolation at finite temperature}

Since the magnetic monopoles are directly observable on the lattice
one can try to check these expectation against the lattice data.
First, however, we should clarify how to distinguish between 
real and virtual particles in the Euclidean formulation of the theory. 
The problem is that in Euclidean space the wave functions of real particles
exponentially decay with time and the difference between virtual and real
particles is not obvious.

Consider first free particles. The answer to our question can be found~\cite{chernodub}
then in closed form within so called polymer representation of field 
theory, see, e.g. Ref.~\refcite{ambjorn}.
One starts with classical action of a particle of mass $M$,
\begin{equation}\label{classical}
S_{{cl}}~=~M\cdot L\,,
\end{equation}
where $L$ is the length of trajectory. 
The Feynman propagator is given then by the sum over
all the paths $P_{x,y}$ connecting the points
$x,y$:
\begin{equation}\label{sum}
G(x-y)~\sim~\sum_{{P_{x,y}}}\exp (-S_{{cl}}\{P_{x,y}\})\,.
\end{equation}
To enumerate all the paths one uses latticized space and the sum (\ref{sum})
with classical action (\ref{classical}) can be evaluated exactly \cite{ambjorn}.
The result is that $G(x-y)$ is indeed proportional to
the propagator of a scalar
particle with mass $m$ related to the original
mass parameter $M$ (see (\ref{classical})) in the following way:
\begin{equation}
m^{2}~\approx~\frac{\mbox{const}}{a}\cdot\big(M(a)~-~\frac{\mbox{const}^{'}}{a}\big)\,,
\end{equation}
where the constants are known for a particular lattice regularization.
Note that for the physical mass, $m^{2}$ be independent of $a$
the original mass parameter  M(a)
is to  be tuned  to
$\mathrm{const}^{'}/a$.

Finite temperature $T$ is introduced through compactification of the time
direction into a circle of length $1/T$ so that the
points 
$$x~=~({\bf x},x_{4}+s/T)$$
where $s$ is an integer number, are identified. 
The physical meaning of $s$ is that it counts
the number of wrapping in time direction.
Intuitively, it is quite clear that the wrapped trajectories
in the Euclidean space  encode
information on real particles at $T\neq 0$ in Minkowski space.
Indeed, compactness of the time direction means that the wrapped
trajectories correspond to particles which exist 'for ever', that is
real particles.

Formally, the proof runs as follows \cite{chernodub}.
One introduces Fourier transform of trajectories wrapped $s$ times:
$$
G_{s}({\bf p})~=~\int d^{3}x \, G({\bf x},t=s/T) \, \exp(-i{\bf px})\,.$$
The $s=0$ case corresponds to unwrapped trajectories which are 
independent of temperature, $G_{0}\equiv G_{\mathrm{vac}}$.
Then one can derive equation:
\begin{equation}\label{bose}
\frac{1}{e^{{\omega_{{\bf p}}/T}}~-~1} = \frac{\sum_{s} G_{s}}{2\ G_{0}}\,,
\end{equation}
which explicitly realizes our intuition by relating the sum over wrapped 
trajectories to the Bose-Einstein distribution.

Further refinement on (\ref{bose}) are possible. First, normalization on
the propagator at $T=0$,
$$G_{0}~=~4/(a^{2}\omega_{\bf p})$$ is awkward since 
$G_{0}$ depends explicitly on the lattice spacing $a$.
Also, we expect that the properties of the monopoles are actually
constrained by environment. To at least partly account for the
interaction with the environment we introduce non-zero
chemical potential $\mu$ and effective number of degrees of freedom
of the monopoles, $N_{df}$.
One can show then that the density of real particles,
\begin{equation}
\rho(T) = \int \frac{d^{3}{\bf p}}{(2\pi)^{3}} \frac{N_{df}}{e^{(\omega_{{\bf p}}+\mu)/T} -1}\,,
\end{equation}
equals to the average number of wrapping per unit 3d volume,
\begin{equation}\label{central}
\rho(T)~=~n_{wr}~=~\langle|s|\rangle/V_{3d}\,.
\end{equation}

Let us  mention the simplest case of static trajectories.
The static trajectories look the same in
Euclidean and Minkowski spaces and obviously
correspond to real particles. Thus, Eq. (\ref{central})
is true in case of static trajectories, also if the
particles interact strongly. At first sight, the static case
 is quite an academic limit. In fact, it is not.
The point is that one can argue on general grounds
that the whole of non-perturbative physics at high temperatures
reduces to  magnetostatics \cite{ginsparg} with strong interaction.
Moreover, detailed lattice studies \cite{ishiguro}
reveal that the approximation of static trajectories works well
for monopoles already around $T=2.4 T_c$ where $T_c$
is the temperature of the deconfinement phase transition.

We will accept (\ref{central}) as definition of density  of real particles 
for lattice studies of the monopole trajectories.

\subsection{Transition chain:~condensate $\to$ liquid $\to$ gas}

Lattice data on the wrapped trajectories turn in fact exciting.
First of all, the change of character of the trajectories near
the phase transition has been noticed in many papers.
Namely,  the infinite, or percolating cluster disappears.
Thus, no tachyonic mode at $T>T_c$, as is expected.

Even more remarkable, the long monopole trajectories 
tend to become time-oriented. As we have just discussed, time-oriented
trajectories correspond to real particles in thermal plasma. 
Moreover, remember that monopoles are just marking
magnetic strings, which is the primary object\footnote{Note that presence of monopoles in thermal plasma
was speculated also in recent papers \cite{chris}.  
Monopoles considered in these papers are essentially Dirac monopoles
and it is not clear how to reconcile such models with the
well known observation that there is no local field theory which
could accommodate both magnetic and electric charges. 
There is no such no-go theorem for the 'monopoles' considered
here since they are rather building blocks of the magnetic strings
than particles, see discussion above.
Moreover, there is no relation between the monopoles introduced
in Ref.~\refcite{chris} at non-zero temperature and  condensate
at zero temperature  while such a relation is a guiding principle
in paper~\refcite{chernodub}.}.
We just follow monopoles because the equations, like (\ref{central})
can be derived explicitly in case of particles. However,
as far as the qualitative behavior is concerned, direct
observations on strings (that is, surfaces) reveal the same
qualitative change: above $T_c$ the strings percolate
only in space--,  not time-- directions, for discussion and
references see Ref.~\refcite{greensite}.

One can conclude that, qualitatively, there is no doubt that 
around $T=T_c$ there is transition of part of the 
tachyonic mode into physical degrees of freedom
of the Yang-Mills plasma. The reality of the magnetic component of the plasma seems to be confirmed. However, it is only quantitative analysis
that cannot tell, how significant this component is.
 
Quantitative analysis \cite{chernodub} of the lattice data on wrapped trajectories
\cite{ejiri} reveals that dependence of the density of monopoles
in plasma is  different in two ranges of temperatures.
First, at temperatures
$$T < T < 2\,T_c$$
the density of thermal monopoles is approximately constant:
\begin{equation}\label{constant}
\rho(T)~\approx~T_c^{3}\,.
\end{equation}
It is interesting to compare this number with density
of ultrarelativistic non-interacting particles at such temperatures.
At $T=T_c$ the density (\ref{density}) is about 10 times higher than the density
of the ideal gas (with one degree of freedom). 

Thus, at temperatures just above the phase transition the monopoles are quite dense and, 
moreover, their density does not depend on the temperature, similarly to a liquid.

Starting from $T\approx 2T_c$ the density of thermal monopoles
begins to grow and at large $T$ the density can be approximated as
\begin{equation}\label{gas}
\rho(T)~\approx~(0.25 T)^{3} \qquad (T>2T_c)\,,
\end{equation}
where we keep only the term dominating at high temperatures.
If we compare now (\ref{gas}) to the ideal-gas case
we find that at very high temperatures the monopoles correspond to much less than
one degree of freedom. 

Thus, there is a chain of transformation of
the magnetic degrees of freedom, as function of
temperature. First, there is quantum condensate.  Then, it melts
into a liquid. And at higher temperatures the liquid is evaporated into a gas.

Note an amusing analogy with physics of superfluidity \cite{landau}.
In case of liquid helium there exist a superfluid and ordinary 
components of liquid. With increasing temperature
the superfluid component undergoes transition into ordinary
liquid. This is analogy of the deconfinement phase transition 
which is associated with vanishing of the quantum condensate and
release of magnetic degrees of freedom into plasma. In terms
of trajectories, infinite cluster existing at zero temperature
is transformed into time oriented trajectories (or, respectively,
surfaces in the string language). At higher temperatures,
$T>2T_c$ the liquid is evaporated into gluon gas. One can expect
that the properties of plasma are described by perturbative physics
with better and better accuracy.

\subsection{Resolving mysteries of phenomenology?}

The observation (\ref{constant}) that the density of the magnetic 
component of plasma is approximately a constant,
$$
\rho_{\mathrm{magnetic}}  \approx {\mathrm{const}}\,,  
\qquad 
T_c~<T~<~2T_c\,,
$$
 might have remarkable
phenomenological implications. The point is that the properties of the
plasma, as are observed at RHIC~\cite{shuryak},
indicate that near the phase transition the plasma is rather a liquid.
 Such a conclusion follows,
in particular, from evaluation of the viscosity which turns to be small\cite{teaney}. 
However, up to now there has been no key to answer
the question, what is dynamics behind these plasma properties.
Lattice measurements which imply (\ref{constant}) might provide a key
to identify particular component of the plasma which looks as
a liquid. Of course much more should be done to be sure of this conclusion.

\section{Phase transition as change of dimensionality }

\subsection{Two types of strings, in 4d and 3d}

In Sec.~2.2. we considered percolation at finite temperature in some
detail. In particular, we argued that wrapped trajectories 
in Euclidean space correspond 
to real (thermal) particles in the Minkowski space. 
The case we considered is in fact Higgs theory of a single 
complex field.
Now we would like
to emphasize that there is deep difference between the Yang-Mills case
and the $\phi^{4}$ case. Namely, in the Yang-Mills case the deconfinement
phase transition is in some sense a change from four- to three-dimensional 
physics. The difference between the two theories  can be readily understood
in simple geometrical terms if we use the strings language for the Yang-Mills
case.

To begin with, we argued above
that in the deconfinement phase 
both electric and magnetic strings have vanishing tension:
\begin{equation}\label{both}
\sigma_{\mathrm{electric}}~=~\sigma_{\mathrm{magnetic}}~=~0\,,
\end{equation}
which implies percolation of two types of strings. 
In four dimensions, however, such percolation is
 hardly possible since each string is a 2d object and
 two types of percolating strings
 would occupy the whole space:
$$2d~+~2d~=~4d~.$$
True, the situation is rather marginal  since strings are expected
to intersect along a submanifold   of dimension $d=0$ and whether it is possible
to avoid intersections depends on details
of the problem.

From the classical paper \cite{thooft} we learn that under feasible assumptions
(like existence of mass gap) 
the resolution of the uncertainty is that percolation of both types of strings 
in 4d is not possible. Indeed in the language of the strings the result of 
Ref.~\refcite{thooft} is that either
$$
\sigma_{\mathrm{electric}}~\neq ~0~, \qquad \sigma_{\mathrm{magnetic}}~=~0\,,$$
or vice verse.

How it is possible then to realize (\ref{both}) at all?
The answer is that the price is violation of O(4) rotational invariance.
The strings should percolate differently  in time and in space directions.
At finite temperature this does not contradict any general principle
since it is only the time coordinate that is compactified.
Thus, at $T\neq 0$ one can consider separately temporal and spatial
Wilson loops\footnote{In fact, instead of the temporal Wilson loop 
one should introduce correlator of two Polyakov lines at large spatial distances.
For recent discussion see, e.g., Ref.~\refcite{andreevzakharov}.},
and temporal and spatial 't Hooft loops. 

Here we come to one of central observations concerning strings in 
4d and 3d spaces. Namely, strings in 3d (or in a time slice of 4d space)
are loops (trajectories)\footnote{We do not discuss here finite thickness
of strings.}, and for two types of strings
$$1d~+~1d~=2d~<3d\,,$$
so that the strings can percolate simultaneously.

For the string tension associated with the temporal loops we get from
our basic relation (\ref{both}):
\begin{equation}\label{both1}
\sigma_{\mathrm{electric}}^{\mathrm{temporal}}~=~\sigma_{\mathrm{magnetic}}^{\mathrm{temporal}}~=~0\,,
\qquad T~>~T_c\,.
\end{equation}
Thus, zero tension have strings which can be open on the lines parallel to
the time direction. As we explained above, vanishing tension means 
percolation. To ensure (\ref{both1}) we need both electric and magnetic strings
percolate in directions perpendicular to the time axis, i.e. in 3d space\footnote{For discussion of 
geometry of strings, their percolation at 
finite temperature and corresponding lattice data, in  particular, Ref.~\refcite{deforcrand}.}.

Next, one can readily understand that at $T>T_c$ both spatial Wilson
and 't Hooft loops exhibit area laws:
\begin{equation}\label{both2}
\sigma_{\mathrm{electric}}^{\mathrm{spatial}}~\neq~0\,,
\qquad
\sigma_{\mathrm{magnetic}}^{\mathrm{spatial}}~\neq~0\,,
\qquad 
T > T_c\,.
\end{equation}
Unlike the 4d case, Eq. (\ref{both2}) is consistent with percolation
of strings in 3d because in 3d there is no representation like
(\ref{open}). On the other hand, percolation of magnetic strings
(1d objects) still disorders the spatial Wilson line and percolation
of electric strings disorders the 't Hooft line. Hence, Eqs. (\ref{both2})
are valid \footnote{For a related discussion see \cite{chernodub1}.}. 
In other words, at $T>T_c$ we expect percolation of both types of strings in 3d.
From the 4d perspective these strings are 2d surfaces nearly parallel to the
$x_{4}$ direction. In the 3d projection they are percolating lines.

\subsection{Three mechanisms under same disguise}

In Sec.~2.2 we emphasized connection between wrapped trajectories
and real particles in thermal plasma. Here, we would like to 
emphasize that we discussed in fact different mechanisms which 
can be revealed by counting the wrapped trajectories.

{\it Percolation of trajectories in 4d}. Consider theory of a 
free complex scalar field in 4d and at temperature $T=0$. 
In the polymer representation
\cite{ambjorn,maxim} the theory is encoded in properties of particle trajectories.
In particular, if  there is no tachyonic mode, $m^{2}>0$, then there
exist only finite clusters of trajectories. The distribution in length $L$
of the clusters is
\begin{equation}
N(L)~\sim~L^{-3}\exp(-m^{2}La)\,.
\end{equation}
Moreover, the 
average radius of the cluster is given by
\begin{equation}
R^{2}~\sim~L\cdot a\,.
\end{equation}
Let us now switch on finite temperature.
Extension of the cluster in the time direction cannot then
exceed $R\sim1/T$. Which means that some of finite clusters
become wrapped trajectories. Counting these trajectories,
one can reproduce the Planck distribution, see (\ref{bose}).

{\it String percolation: from 4d to 3d.} Another mechanism was discussed
in Sec.~3.1. Here we argued that at $T>T_c$ both magnetic and
electric strings percolate only in spatial directions. In other words, 
the long trajectories become time-oriented. 
Let us emphasize that this is absolutely different mechanism
than simple percolation. In case of percolation, the properties
of trajectories are not changed by switching on temperature.
Just some long trajectories are 'caught' by the periodicity
condition at $x_{4}=s/T$ and become wrapped trajectories.

In case of the ``change of dimensionality'' of string percolation,
the trajectories become time-oriented. The effect
is not simply periodicity imposed on the boundary.
How one can understand such a non-local effect?
An explanation is suggested by  AdS/QCD correspondence (for references see, 
e.g., Ref.~\refcite{vz}).
The basic idea is that the tension of strings (both electric and magnetic)
depends on their length. In particular, $\sigma_{\mathrm{magnetic}}=0$
only if the string is long enough, $l\sim\Lambda_{\mathrm{QCD}}^{-1}$.
And only tensionless strings percolate. If one imposes
periodic boundary condition at $x_{4}=s/T$ then 
at the critical temperature the percolating string is no
longer long enough, and
there can be no percolation in the time direction. There is a 
non-local change of properties of strings because the condition
for percolation is non-local: string is to be long enough. 
There is no such condition in case of the standard percolation.
Since percolation in the time direction is no longer allowed
 time-oriented strings become energetically favorable. 

At $T=T_c$ not only the magnetic string changes its tension.
For example, the string tension associated with the spatial
Wilson loop in the model \cite{andreev}
depends on temperature in the following way \cite{andreevzakharov}:
\begin{equation}\label{andreev}
\sigma_{\mathrm{electric}}^{\mathrm{spatial}}~\approx~\frac{T^{2}}{T_c^{2}}
\exp\bigg(\frac{T_{c}^{2}}{T^{2}} -1\bigg)\,, \qquad T>T_c
\end{equation}
Probably, the very position of the phase transition, $T_c$
could be found from condition of self-consistency
of predictions for various strings. But this issue has not been addressed
in literature so far.

{\it Dimensional reduction.} Non-perturbative nature of magnetostatics 
in the Yang-Mills case was discovered first within the scheme
of dimensional reduction which should work at high temperatures,
when the running coupling $g^{2}(2\pi T)$ is small
\cite{ginsparg}. The picture is
quite similar to the case just discussed
with even more asymmetry between the spatial and time
directions. For example, instead of time-oriented
oriented trajectories one would discuss strictly static trajectories.

\subsection{Identifying the mechanism}

Using lattice data on wrapped trajectories and various 
string tensions one can in fact distinguish between various
mechanisms summarized in Sec.~3.2.

In particular, in case of standard, or particle percolation
the length of the wrapped trajectories would be divergent
in the continuum limit, $a\to 0$:
\begin{equation}\label{spacings}
\langle L_{\mathrm{wrapped}}\rangle~\sim~T^{-2}a^{-1}\,.
\end{equation}
Such dependence on the lattice spacing is typical of
the random walk.

On the other hand, the string model with running string tension
is non-local and it allows for a change of character of trajectories
through the whole lattice. The relation (\ref{spacings})
is not valid any longer. Also, there is natural scale for $T_c$
which is position of the horizon in the AdS/QCD schemes,
see, e.g., Ref.~\refcite{andreevzakharov}.

The lattice data on the wrapped trajectories which we discussed
in Sec.~2 clearly favor the non-local model, or string percolation.

Next, let us discuss how to distinguish a general scheme
of ``change of dimensionality'' from the standard dimensional
reduction. A key point here is that the dimensional reduction
is derived within perturbation theory. The perturbation theory,
in turn, can be applied to evaluate $\sigma_{\mathrm{electric}}^{\mathrm{spatial}}$
and $\sigma_{\mathrm{magnetic}}^{\mathrm{spatial}}$, see, e.g., 
Ref.~\refcite{perturbationtheory}
and references therein.
There is ample evidence that at temperatures near 
the phase transition, $T<2T_c$ perturbation theory does not work.
%Moreover, non-perturbative treatments, like AdS/QCD correspondence
%are most successful in the same range $T<2T_c$. 
Thus, dimensional reduction seems not to apply at temperatures
crucial for the ``magnetic liquid'' discussed in Sec.~2. 

As a further potential check of the string mechanism let us mention
possible role of strings in the anomaly. The point is that strings
naturally contribute to the temperature-dependent conformal anomaly:
\begin{equation}\label{estimate}
\langle G^{2}(T)\rangle~\sim~\Lambda_{\mathrm{QCD}}^{2}\,T^{2}\,,
\end{equation}
where $\langle G^{2}(T)\rangle$ is the temperature dependent part of the
gluon condensate which can be studied directly on the lattice,
for review see, e.g., Ref.~\refcite{pisarski}.

Indeed, in time-slice (or in 3d) strings percolate as trajectories,
or loops. For the QCD-related strings which we are discussing
the total length of trajectories is of order
\begin{equation}
L_{\mathrm{total}}~\sim~\Lambda_{\mathrm{QCD}}^{2}V_{3d}\,.
\end{equation}
The strings are time-oriented and extend in the time direction
at distance of order $T^{-1}$. The tension for such strings
(modulo possible 'threshold effects' near $T_c$) is
of order $\sigma\sim T^{2}$, see, e.g., (\ref{andreev}). In this way
we come to the estimate (\ref{estimate})  for the action density
associated with the strings.

\section{Conclusions}

There is amusing possibility that magnetic degrees of freedom
which condense and tachyonic (unphysical) at zero temperature become
a component of thermal Yang-Mills plasma. There are first
indications from lattice simulations that this is indeed the case
but much more is to be done to really confirm the picture.

\section*{Acknowledgments}
We are thankful to O. Andreev, V. Bornyakov, A. Nakamura, M.~I. Polikarpov,
T. Suzuki for useful discussions and to B. Shklovskii and  A.~I. Vainshtein
for communications. 
The work of M.N.Ch. was supported by the JSPS grant No. L-06514
and by Grant-in-Aid for Scientific Research by Monbu-kagakusyo, No. 13135216.
V.I.Z. is thankful to the organizers of the SCGT06 workshop, 
and especially to Prof. K.  Yamawaki for the invitation and hospitality.

\end{document}